# High-Efficiency Resonant RF Spin Rotator with Broad Phase Space Acceptance for Pulsed Polarized Cold Neutron Beams

P.-N. Seo[1a], L. Barron-Palos[2], J.D. Bowman[1b], T.E. Chupp[3], C. Crawford[4], M. Dabaghyan[5], M. Dawkins[6], S.J. Freedman[7], T. Gentile[8], M.T. Gericke[9], R.C. Gillis[9c], G.L. Greene[4,10], F.W. Hersman[5], G.L. Jones[11], M. Kandes[5], S. Lamoreaux[1d], B. Lauss[7e], M.B. Leuschner[6f], R. Mahurin[4], M. Mason[5g], J. Mei[6], G.S. Mitchell[1h], H. Nann[6], S.A. Page[9], S.I. Penttilä[1b], W.D. Ramsay[9,12], A. Salas Bacci[1], S. Santra[6i], M. Sharma[3], T.B. Smith[13], W.M. Snow[6], W.S. Wilburn[1], H. Zhu[5j]

[1]*Los Alamos National Laboratory, Los Alamos, NM 87545, USA*
[2]*Arizona State University, Tempe, AZ 85287, USA*
[3]*University of Michigan, Ann Arbor, MI 48104, USA*
[4]*Department of Physics University of Tennessee, Knoxville, TN 37996, USA*
[5]*Department of Physics, University of New Hampshire, Durham, NH 03824, USA*
[6]*Department of Physics, Indiana University, Bloomington, IN 47405, USA*
[7]*Department of Physics University of California, Berkeley, CA 94720, USA*
[8]*National Institute of Standards and Technology, Gaithersburg, MD 20899, USA*
[9]*Department of Physics, University of Manitoba, Winnipeg, Manitoba R3T2N2, Canada*
[10]*Oak Ridge National Laboratory, Oak Ridge, TN 37831, USA*
[11]*Department of Physics, Hamilton College, Clinton, NY 13323, USA*
[12]*TRIUMF, 4004 Wesbrook Mall, Vancouver, British Columbia V6T2A3, Canada*
[13]*Department of Physics, University of Dayton, Dayton, OH, 45469, USA*

We have developed a radio-frequency resonant spin rotator to reverse the neutron polarization in a 9.5 cm×9.5 cm pulsed cold neutron beam with high efficiency over a broad range of cold neutron energies and transverse momenta. The effect of the spin reversal by the rotator on the neutron beam phase space is compared qualitatively to RF neutron spin flippers based on adiabatic fast passage. The spin rotator does not change the kinetic energy of the neutrons and leaves the neutron beam phase space unchanged to high precision. We discuss the design of the spin rotator and describe two types of transmission-based neutron spin-flip efficiency measurements where the neutron beam was both polarized and analyzed by optically-polarized $^3$He neutron spin filters. The efficiency of the spin rotator was measured to be 98.0±0.8% on resonance for neutron energies from 3.3 to 18.4 meV over the full phase space of the beam., which possessed transverse beam velocities up to 20 m/sec.. As an example of the application of this device to an experiment we describe the integration of the RF spin rotator into an apparatus to search for the small parity-violating asymmetry $A_\gamma$ in polarized cold neutron capture on para-hydrogen by the NPDGamma collaboration at LANSCE.

---

[a] Corresponding author: Tel:+1+919-513-0319, pilneyo@tunl.duke.edu, Department of Physics, North Carolina State University, Raleigh, NC 27695, USA
[b] Present Address: Oak Ridge National Laboratory, Oak Ridge, TN 37831, USA
[c] Present Address: Department of Physics, Indiana University, Bloomington, IN 47405, USA
[d] Present Address: Department of Physics, Yale University, New Haven, CT 06520, USA
[e] Present Address: Paul Scherrer Institute, CH-5232 Villigen-PSI, Switzerland
[f] Present Address: Procure, Bloomington, IN 47408, USA
[g] Present Address: Texas A&M University, College Station, TX 77843, USA
[h] Present Address: Department of Biomedical Engineering, University of California, Davis, CA 95616, USA
[i] Present Address: Bhabha Atomic Research Center, Trombay, Mumbai 400085, India
[j] Present Address: Canberra, CT, USA

# I. INTRODUCTION

A number of experiments in nuclear and particle physics and in condensed matter using polarized cold neutron beams need devices to reverse the neutron polarization. These devices, usually called spin flippers, should ideally possess high spin-flip efficiency, operate reliably and efficiently over a broad distribution of neutron beam phase space, occupy a compact space, should not introduce any material into the neutron beam, and should maintain compatibility with different types of neutron polarization devices. As the continued increase in the brightness of pulsed spallation neutron sources makes polarized neutron measurements possible with a new level of accuracy, there is also a need to improve and develop neutron spin flippers to accommodate the demands of these new types of experiments by taking advantage of the pulsed nature of the neutron source.

In this paper we describe the theory, design, and detailed characterization of a neutron spin flipper which we term a spin rotator. It is a resonant device based on performing NMR on a polarized neutron beam which moves through an orthogonal combination of static and RF magnetic fields in no DC field gradient. It is an active device in the sense that the amplitude of the RF magnetic field is varied in phase with the pulsed neutron source in such a way that the condition for rotating the neutron spins by $\pi$ radians in the rotator is always met for each subsequent neutron velocity class as it reaches the device. Such a resonant RF spin rotator for neutrons is by no means new: it has been implemented by many groups in the past [1]. What is new in our device is the combination of the breadth of neutron phase space over which the efficiency of the rotator is high, the accuracy and detail of the spin-flip efficiency measurements, and the variety of independent methods used to determine the efficiency. To achieve the required breadth of polarized neutron phase space coverage we used neutron spin filters based on transmission through polarized $^3$He gas both as polarizers and as polarization analyzers, and this is another novel aspect of the work. Both relative transmission measurements and reversal of the $^3$He polarization using adiabatic fast passage (AFP) were employed to determine the spin-flip efficiency. We also used a current-mode $\gamma$-detector array and a parity-violating (n,$\gamma$) reaction as a neutron polarization analyzer as an independent verification of the efficiency measurements.

# II. PRINCIPLES OF OPERATION OF RF NEUTRON SPIN FLIPPERS

Neutron spin flippers based on RF magnetic fields have been commonly used for decades in experiments with polarized slow neutrons. There are two main types of RF flippers: adiabatic spin flippers and resonant spin rotators. The operating principle of both devices is well-known from NMR [2] with small modifications due to the fact that the neutron spins are moving in the lab frame.

Adiabatic RF neutron spin flippers consist of a static magnetic field $B_0$, that changes in magnitude along a length of beam, and a perpendicular RF magnetic field whose magnitude varies with a maximum in the middle of the flipper. The operation of the flipper is best understood in the rotating frame of the neutron spin. If we choose the reference frame rotating at the Larmor frequency set by the static part of the field in the middle of the flipper and the RF frequency to match this Larmor frequency, then in the rotating frame the transverse static field disappears and the longitudinal RF field becomes static. In this frame the neutron spin precesses about the remaining effective magnetic field which, as seen by the neutron, slowly varies from transverse up as the beam enters the flipper to longitudinal in the middle of the flipper and finally transverse down at the exit of the flipper. As long as the precession frequency $\Omega$ of the neutron in this rotating frame about this slowly-varying effective field is fast compared to the rate of change of the direction of the effective field, then the projection of the neutron spin onto the effective field is an adiabatic invariant and the spin follows the effective field with high accuracy. The probability of the depolarization during rotation depends exponentially on the adiabatic parameter $\lambda = (\mu_n \frac{B_0^2}{\hbar}) / (v \frac{dB_0}{dz})$ where $\mu_n$ is the neutron magnetic moment and $v$ the neutron velocity. This adiabatic condition is easy to meet for slow neutron energies and such a spin flipper is relatively easy to realize. Adiabatic RF spin flippers for neutrons were first used for ultra-cold neutrons [3,4] and later also for cold neutron beams [5,6]. The theory for their operation has been further developed to take into account higher-order effects from Berry's phase [7,8]. The spin-flip efficiency can be high for a broad spectrum of neutron energies as long as the neutrons satisfy the adiabatic condition.

Resonant RF spin flippers also employ NMR principles. In this case, the static field is constant and the RF frequency is chosen to match the Larmor precession frequency throughout some region of space. When the neutron with speed $v$ passes through this region and meets the resonance condition, its spin is rotated by an angle of $\pi$ radians. Unlike the adiabatic spin flipper, the spin-flip efficiency of the resonant spin rotator is directly sensitive to the neutron velocity. It therefore possesses high efficiency only for neutrons in a narrow class of velocities.

However, the action of these two types of spin flippers on the phase space of the beam is not quite the same. Since the adiabatic flipper by necessity possesses a static magnetic field gradient, the Stern-Gerlach effect in such a flipper will steer the neutron beam in slightly different directions for the two spin states. In addition, by Maxwell's equations the static magnetic field gradient necessarily leads to field gradients in regions outside of the spin flipper. Finally, since the static magnetic field is different at the two ends of the device the neutron kinetic energy can change. As we discuss below, a resonant RF spin flipper need posses none of these effects.

An example of an experiment which places stringent limits on neutron spin-dependent phase space changes is the NPDGamma experiment, a measurement of the small parity-violating weak $\pi$-nucleon coupling constant, $f_\pi^1$, in the reaction of $\vec{n} + p \rightarrow d + \gamma$. In the experiment polarized cold neutrons are captured by protons in a liquid para-hydrogen target and the $\gamma$-ray asymmetry $A_\gamma$ with respect to the neutron beam polarization is measured [9]. The experiment aims to determine $f_\pi^1$ unambiguously by measuring $A_\gamma$ with $1\times10^{-8}$ precision [10].

At this level of precision, spin-dependent changes in the phase space of the polarized neutron beam must be strictly minimized. The Stern-Gerlach effect that would necessarily be present in an adiabatic RF spin flipper will deflect the neutron beam along the neutron polarization direction. The resulting spin-correlated spatial shift in the neutron capture distribution in the hydrogen target changes the solid angle seen by the γ-detectors and, therefore, leads to a spin-correlated asymmetry in the γ-detector array that is difficult to distinguish from the parity-violating signal that the experiment measures. Although this effect can be minimized by employing a smaller static field gradient in the adiabatic flipper, this requires the length of the flipper to be increased to maintain the adiabatic condition for all of the neutrons in the spectrum and therefore requires all downstream elements of the apparatus to be widened to continue to accept all events from the diverging cold neutron beam. To eliminate this potential source of the systematic uncertainty, we decided to design a RF spin rotator that possesses negligible static field gradients.

We call this device a RF spin rotator (RFSR) as opposed to a RF spin flipper to emphasize that this device does not change the kinetic energy of the neutrons on resonance. The change in potential energy of the neutron inside the RF field region is caused by the exchange of energy of the neutron with the RF field, leaving the kinetic energy unchanged. There are two equivalent ways to see why this spin rotator has this property.

### A. Semi-classical view

Consider a mono-energetic neutron wave packet moving along the $+\hat{y}$-axis starting from -∞ with an initial polarization in the $+\hat{z}$ direction. Let it move in a uniform and static magnetic field $B_0\hat{z}$, and in a uniform and time-dependent magnetic field which rotates normal to the $\hat{z}$-axis with frequency Ω, $\vec{B}(t) = \hat{y}B_1\cos(\Omega t) + \hat{x}B_1\sin(\Omega t)$, over a length $d$ on the $+\hat{y}$-axis. Let the Larmor precession frequency be $\omega_0 \equiv \mu_n B_0/\hbar$. Differences between this model and the real spin rotator are small.

Consider the semi-classical view used in NMR in which one performs a transformation from the inertial rest frame of the neutron into a frame rotating about the $+\hat{z}$-axis with frequency Ω. If $\Omega = \omega_0$ (on resonance), then in the rotating reference frame and in the RF field region the neutron sees only an effective magnetic field $\vec{B}_{\text{eff}} = B_1\hat{y}$ along the $+\hat{y}$-axis which is normal to the initial spin direction. Therefore, in this frame the neutron starts to precess about the $+\hat{y}$-axis and if the time it spends in the RF field is just right then the spin rotates by π radians before exiting the region. Notice that if we also view the motion of the neutron before and after the RF field region in this rotating frame then the neutron never experiences a change in the effective field along the $+\hat{z}$-axis. But this change is what would be required to change the potential energy of the neutron. And in addition the transformation to the rotating frame removes the energy exchange of the neutron with the RF field by turning this into a static field. Therefore $\vec{\mu}_n \cdot \vec{B}_{\text{eff}} = 0$ in the rotating frame on resonance and the kinetic energy of the neutron does not change.

This is not true off-resonance where $\Omega \neq \omega_0$ and the effective magnetic field possesses a component along the $+\hat{z}$-axis when the neutron enters the RF field region. There is a change in potential energy at boundary and, therefore, a change in the neutron kinetic energy. In the rotating frame the effective field makes an angle $\tan\theta = (\omega_0 - \Omega)/\omega_1$ where $\hbar\omega_1 = \mu_n B_1$ and if the neutron spends the same amount of time in the RF field then it precesses to an angle $\pi/(2-2\theta)$ and the change in its potential energy as it transverse the boundary of the RF region, is therefore $\hbar\Omega\cos\pi/(2-2\theta) \approx \hbar(\omega_0 - \Omega)$ for small values of $\theta$.

## B. Quantum-mechanical view

The same result is obtained using a quantum mechanical approach. One can solve for the time-dependent wave functions for the spin up and spin down components by the methods of one-dimensional quantum mechanics [11,12]. The derivation assumes a wave function of the form for a plane wave traveling in the $+\hat{y}$ direction

$$\psi^\pm(y,t) = \begin{bmatrix} A_n^\pm \\ B_n^\pm \end{bmatrix} e^{ik_n y} e^{-i(\omega_n \pm \Omega/2)t},$$

here $n$ indexes the three regions of space; before, inside, and after the time-dependent field region located in $0 \leq y \leq d$, and +/- refers to up/down spin states. Then one matches the wave functions and their derivatives in the usual way at two boundaries. This procedure generates 16 different components to the time-dependent wave function in the RF field region. For each of the two spin states and for each of the reflected and transmitted waves, there are four independent components of the wave function. Two components correspond to the two different momenta in the dispersion relation generated by substituting $\psi^\pm(y,t)$ into the Schrödinger equation and for the case of no energy exchange with the time-dependent field. In the region before the RF field these momenta are $k_i^+ = \sqrt{\omega_{kin} - \omega_0}$ for spin up and $k_i^- = \sqrt{\omega_{kin} + \omega_0}$ for spin down, $k_i = \sqrt{\omega_{kin}}$ represents the wave number ($\omega_{kin}$ kinetic energy) of the neutron in vacuum; no static or RF fields present. To simplify derivation, units $\hbar = 2m = 1$, where $m$ is the neutron mass, are used. The other two components correspond to these two momenta with energy exchange with the RF field. Under the usual assumptions that the time-dependent magnetic field is weak enough that only one RF photon is exchanged between the neutron and the RF field, and that the kinetic energy is large compared to the potential energy of the magnetic moment in the fields and also to the energy exchanged with the time-dependent field, one can solve for the amplitudes for the neutron spin to be up or down in the region past the time-dependent field $y \geq d$ assuming it entered with spin up. Using the notation of reference [11], the transmitted wave function becomes

$$\xi^\pm(y,t) = \begin{bmatrix} T_0^+ e^{-i\omega_0 y/v} \\ T_1^- e^{-i(\Omega - \omega_0)y/v} e^{i\Omega t} \end{bmatrix} e^{i(k_i y - \omega_{kin} t)}$$

with $T_0^+ = e^{-i\varepsilon d/v}[\cos(\omega_A d/v) + i(\varepsilon/\omega_A)\sin(\omega_A d/v)]$ and

$T_1^- = -i(\omega_1/\omega_A)e^{i\varepsilon d/v}\sin(\omega_A d/v)$ . Here $\omega_A^2 = \omega_1^2 + \varepsilon^2$, $\varepsilon = \Omega/2 - \omega_0$, and $\omega_1 \ll \omega_0$. Then, the spin-flip probability is given by

$$f = |T_1^-|^2 = \frac{\omega_1^2}{\omega_A^2}\sin^2(\omega_A d/v). \tag{2.1}$$

On resonance $\varepsilon=0$ and for $\omega_1 d/v = \pi/2$, $f$ is unity. For the final spin-down state we can read off the momentum and energy from the wave function: $k^- = k_i^+ - \omega_0/v$ and total energy $\omega = \omega_{\text{kin}} - 2\omega_0$. Neglecting quadratic term of the Zeeman energy, the kinetic energy is $k^{-2} \cong k_i^{+2} - 2\omega_0 k_i^+/v = \omega_{\text{kin}} - \omega_0$. Comparing with the initial kinetic energy of the spin-up neutron before entering the RF field, $k_i^{+2} = \omega_{\text{kin}} - \omega_0$, we see that indeed the kinetic energy of the neutron has not changed in agreement with the conclusion drawn using the rotating reference frame.

Off resonance the momentum and total energy are $k^- = k_i^+ - (2\varepsilon + \omega_0)/v$ and $\omega^- = \omega_{\text{kin}} - \Omega$ . The kinetic energy is $k^{-2} \cong k_i^{+2} - 4\varepsilon k_i^+/v - 2\omega_0 k_i^+/v = \omega_{\text{kin}} - \omega_0 - 2\varepsilon$ and a change in kinetic energy is $-2\varepsilon$ as determined above. The conclusion of this derivation is the same: on resonance where $\varepsilon=0$, the neutron kinetic energy is not changed by the interaction with the RF field.

How does the realistic RFSR device deviate from this idealized case? First, the RF field direction does not rotate in space as assumed but oscillates in magnitude. As usual one can decompose the RF field as a sum of a rotating and a counter-rotating component, and if one is close to resonance then the other is far from resonance and is negligible for the small amplitudes of the RF fields employed in this work. Second, the amplitude of the RF field is in fact not uniform inside the spin rotator. Instead it has a spatial distribution along the trajectory of the neutron determined by the boundary conditions of the RF field inside the aluminum shield. However, it is clear that on resonance this non-uniformity still does not lead to a change in the kinetic energy of the neutrons to first order because the effective field, although now time-dependent, still satisfies $\vec{\mu}_n \cdot \vec{B}_{\text{eff}} = 0$ in the rotating reference frame. Therefore even a realistic RF spin rotator is expected to leave kinetic energy of the neutron beam unchanged to high accuracy.

## III. DESCRIPTION OF THE APPARATUS

Figure 1 shows a schematic of the NPDGamma experiment and setup for the measurement of the spin-flip efficiency in the pulsed cold neutron beam line flight path 12 (1FP12) at the Los Alamos Neutron Science Center (LANSCE). Neutrons from the 20 Hz spallation target are moderated by the 1FP12 partially-coupled liquid hydrogen moderator [13] and delivered by a 20.7 m long supermirror ($m$=3) neutron guide with cross sectional area of 9.5 cm × 9.5 cm to a magnetic and radiological shielding structure which houses the experiment. A $m$=3 supermirror neutron guide transmits neutrons with transverse velocities <20 m/sec with high efficiency. To prevent frame overlap of the neutrons from different pulses which would degrade the one-to-one correlation of neutron energy with a time-of-flight the neutrons with energy less than 1 meV are absorbed by a frame-definition chopper located between the moderator and the experiment [14].

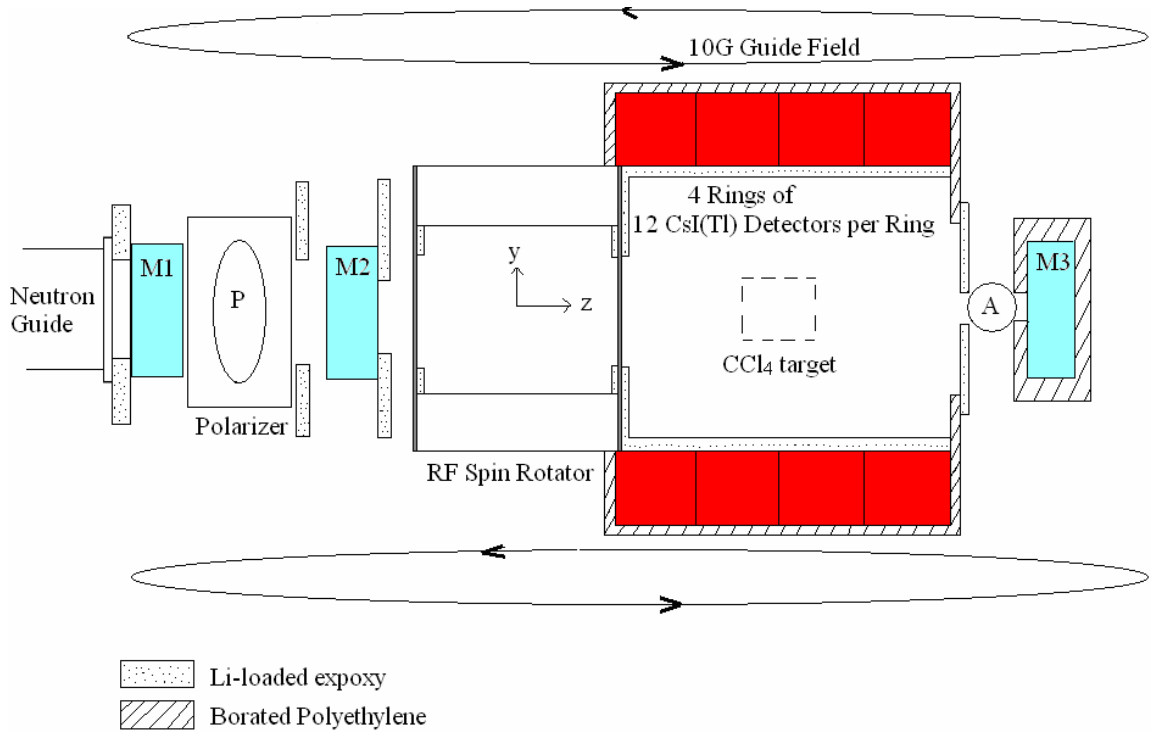

FIG 1. A schematic side view (not to scale) of the NPDGamma apparatus and setup for the spin-flip efficiency measurement on 1FP12 at LANSCE. M1, M2, and M3 are $^3$He ion chambers. The polarized $^3$He polarizer cell P is placed between M1 and M2 and the polarized $^3$He analyzer cell A is mounted in the front of M3 which is 1.65 m downstream from the exit window of the neutron guide. The analyzer cell and M3 are used for measurement of the spin-flip efficiency and to monitor the polarization of the beam passing through the liquid parahydrogen target. Beam collimators are made from $^6$Li-loaded epoxy material and neutron shielding from borated polyethylene. The entire apparatus is in a uniform 1 mT magnetic field.

Neutron beam monitors M1, M2 and M3 are parallel plate $^3$He ionization chambers filled with an admixture of $^3$He, $^4$He, and $N_2$ to a total pressure of one atmosphere [15]. The stable and linear neutron beam monitor M1 is used to normalize the neutron beam intensity pulse-by-pulse with 1% accuracy. Monitors M1 and M2 are used for the beam polarization measurement. A monitor pair M2 and M3 is used to follow the ortho-para ratio of the liquid para-hydrogen target of the NPDGamma experiment (the target is not shown in Fig. 1) [16], to study the depolarization of the neutron beam in the hydrogen target, and to measure the spin-flip efficiency of the RF spin rotator. The monitors M1 and M2 each absorb 4% of the beam at 4 meV and the M3 absorbs all the incident cold neutrons.

For the spin-flip efficiency measurement the beam is collimated to 10 cm diameter using $^6$Li-loaded epoxy material (9% by weight, 90% enrichment of $^6$Li). After M1 the unpolarized neutrons enter the spin filter, a cell of polarized $^3$He gas [17]. Because of the large spin-dependent neutron-$^3$He capture cross section, neutrons with spin parallel to the $^3$He spin, the triplet state, are transmitted without significant attenuation, while those with antiparallel spin are absorbed. The $^3$He cell of 12 cm in diameter can polarize the entire cold neutron beam phase space, and the magnitude of the polarization variation with neutron transverse momentum is of order <~0.1%. The spin filter is operated in a 1 mT homogeneous ($\Delta B_0/\Delta r$<10 µT/m) vertical magnetic field. The polarized beam passes through a second 10 cm collimator before entering M2. For the measurement a third 10 cm diameter collimator is inserted on the downstream side of M2. The neutron beam polarization is calculated from relative neutron transmission intensities of the monitors M1 and M2 [18].

A resonant RFSR is used for neutron spin reversal at 20 Hz. In the NPDGamma experiment after the RFSR the polarized cold neutrons enter a liquid parahydrogen target (LH$_2$), where they are captured by protons. Deuterons from the capture reaction decay to the ground state by emitting 2.2-MeV γ-rays that are detected by an array of 48 CsI(Tl) detectors [19] viewed by vacuum photodiodes coupled to low-noise solid-state electronics [20]. The neutron polarization analyzer, a polarized $^3$He cell in front of M3, is used to analyze the polarization of the beam that passes through the hydrogen target or the performance of the RFSR. Due to the small size of the spherical analyzer cell, 2.5 cm in diameter, the beam entering the analyzer is collimated to 1.3 cm diameter. M3 is shielded by borated polyethylene against background neutrons.

The neutron kinetic energy will change as discussed above if the static magnetic field changes. To minimize this effect and provide conditions for the resonant RF spin reversal the entire experiment is immersed in a 1 mT magnetic field. The homogeneity requirement of the field is set by the need to suppress a possible source of a false asymmetry caused by the up-down Stern-Gerlach steering of the polarized beam after the RFSR. An additional requirement is that the direction of the field respect to the γ-detector in the location of the LH$_2$ target has to be known with the accuracy of 20 mrad to prevent the parity-conserving left-right asymmetry to have a vertical component and thus to produce a false up-down asymmetry in the experiment. Measured field orientation with respect to the detector is less than 20 mrad, which is the gradient requirement.

In the NPDGamma experiment the γ-ray asymmetry $A_\gamma$ is determined from γ-ray yields obtained by the upper half and the lower half of the γ-detector array for every neutron pulse. The γ-asymmetry in 1FP12 is measured with the accuracy of about $10^{-3}$ per neutron pulse with 100-µA average proton current on the spallation target. Since it is not practical to require the efficiencies of the detector halves to be equal to this precision, the spin rotator is needed to reverse the sign of the asymmetry with respect to the γ-detector array.

The main handle to control systematic uncertainties in the NPDGamma experiment is therefore the frequent reversal of the neutron spin direction. The neutron polarization can be reversed with two independent spin flips without changing the static magnetic field. The polarization direction of the $^3$He in the spin filter can be reversed with the adiabatic fast passage (AFP) technique. The frequent use of this spin flip is limited by the time of a few seconds it takes to perform it and a slow recovery of a small $^3$He polarization loss per AFP flip. The RFSR is designed to reverse the neutron spin direction at frequencies from zero to 60 Hz and was chosen to match the 20 Hz pulse frequency of the LANSCE pulsed spallation source. Although a third neutron spin flip can in principle be performed by reversing the direction of the vertical 1 mT static field, a potential slow relaxation of the magnetization in the steel of the magnetic shielding precludes this method at least in practice in 1FP12.

In the LANSCE spallation source neutrons are produced by short, a few hundred nanoseconds long, proton pulses which strike two tungsten targets. The moderated neutrons have an energy distribution and therefore they have different arrival times in the experiment.

Before entering the RFSR the neutron spin precesses in the static field of $\vec{B}_0 \hat{y}$ where $B_0$=10 mT, see Fig. 1 for the coordinates. Inside the RFSR the static and time dependent magnetic fields are present and in the rotating reference frame the spin rotates with angular frequency of $\omega_1 = \frac{\mu_n B_1}{\hbar}$ about the effective magnetic field $\vec{B}_{\text{eff}} = (B_0 - \frac{\Omega}{\mu_n}\hbar)\hat{y} + B_1\hat{z}$ which on resonance is $\vec{B}_{\text{eff}} = B_1\hat{z}$ and the neutron spin rotates in the $\hat{x}\hat{y}$-plane. To rotate the incoming spin by $\pi$ radians compared to incoming to the outgoing, the spin has to remain in the RF field for a time $t = n\pi\hbar/\mu_n B_1$, where $n$=1,3,5,…. Since the length $L$ of the spin rotator is fixed, the time depends on neutron velocity $v$. If the time is $t = L/v$, then the condition for the $\pi$-radian rotation is satisfied. To achieve a high spin-flip efficiency for each neutron velocity, the amplitude $B_1$ has to be

$$B_1 = \frac{n\pi\hbar}{\mu_n}\left(\frac{l}{L}\right)\frac{1}{t_{tof}}, \tag{3.1}$$

where $l$ is the distance from the moderator to the spin rotator (22 m in the NPDGamma experiment on 1FP12) and $t_{tof}$ is the neutron time-of-flight. A result of Eq. (3.1) is that the RF field strength $B_1$ has to be varied during a neutron pulse as $1/t_{tof}$.

According to Eq. (2.1) the spin-flip probability for neutron with velocity $v$ depends upon $B_1$ and $(\Omega - \omega_0)$ since time $t$ that the neutron with energy $E_n$ spends in the RFSR, is fixed. In the real RFSR $B_1$ is not constant on different neutron trajectories because the field is produced by a short solenoid. Therefore, $(\Omega - \omega_0) \neq 0$ can be a consequence of an inhomogeneous RF magnetic field across the RFSR or of $\Omega$ not being perfectly tuned to resonance.

Figure 2 shows the spin-flip probability calculated using Eq.(2.1) for the 30 cm long RFSR solenoid as a function of $(\Omega - \omega_0)$ for three different neutron energies; 1 meV (solid line), 4 meV (dotted line), and 10 meV (dashed line). The RF amplitude $B_1$ is selected so that on resonance the neutron spin rotates by $\pi$ radians. In the 1 mT static field the 10-meV neutron performs eight Larmor precessions in the RF field whereas the spin of a 1 meV neutron performs 30 precessions. A 2-µT change in the static field moves the spin out of resonance and reduces the spin-flip probability for 1 meV neutron by 4.5%, for 4 meV neutron by 1.1%, and for 10 meV neutron by 0.5%. A 5% change in the optimum strength of $B_1$ reduces the spin-flip probability by 0.7% on resonance.

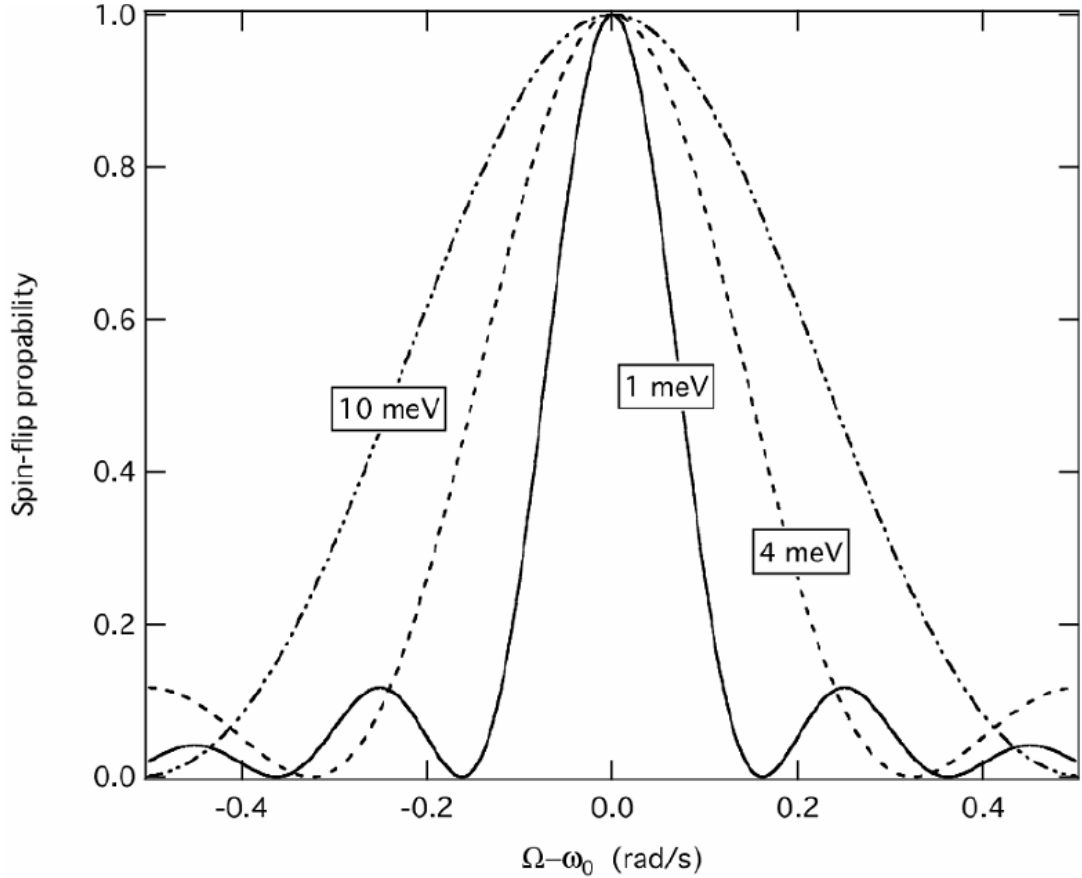

FIG 2. Spin-flip probability as a function of $(\Omega-\omega_0)$ for the RFSR of the NPDGamma experiment on 1FP12 at LANSCE for three different neutron energies. At each energy the amplitude of RF field $B_1$ has been selected so that a full spin flip takes place on resonance.

## IV. DESIGN AND CHARACTERIZATION OF THE RF RESONANT SPIN ROTATOR

### A. RF magnetic field design for the RF resonant spin rotator

The RF field of the RFSR is produced by a solenoid inside an aluminum shield and can be driven at Larmor frequency of approximately 29 kHz for the 10 mT guide field. The amplitude $B_1$ is not constant inside the spin rotator since it is produced by a short solenoid with a finite radius and it is modified by the boundary conditions from the aluminum shielding. A more appropriate design parameter is the integral of the amplitude of the RF field along the neutron trajectory over the length of the spin rotator,

$$\int_{-L/2}^{L/2} B_1(r,z)\,dz,$$

where $r$ is the radial distance from the axis. This integral is directly proportional to the number of precessions of the neutron polarization along the effective magnetic field in the rotating frame. In order to achieve high spin flip efficiency the neutron spin has to perform several Larmor precessions about the effective field in the rotating frame as it passes through the spin rotator.

The dimensions of the solenoid and its aluminum RF shielding enclosure have been optimized for the neutron beam phase space available on FP12 at LANSCE. The spin rotator can accept a 10 cm x 10 cm beam area. The length of the solenoid is defined by the RF amplitude $B_1$ and the adiabatic requirement. These factors limited the maximum neutron energy which can be reversed by our RFSR to ~100 meV. The diameter was optimized for high spin-flip efficiency resulting in a length-to-diameter ratio of 1:1. The aluminum shielding dimensions must be larger than the RF solenoid to reduce the transverse gradients at the position of the entrance and exit windows, which could lead to substantial depolarization of the neutrons on off-axis trajectories. The RFSR must also possess efficient electromagnetic shielding to prevent RF power from reaching the rest of the experiment and potentially causing a false systematic effect.

To satisfy these design criteria we chose a RF solenoid of $S$=15 cm in radius and $M$=30 cm in length with its axis along the neutron beam. The solenoid has a single layer of 273 turns of 18 gauge Cu wire wound on a nylon spool. The coils are shielded with an Al cylinder (R=20 cm, L=40 cm) with Al caps at the ends. The thickness of the wall of this shielding is 5 mm and the entry and exit windows are 0.5 mm thick, which is the skin depth of the 29 kHz RF field in Al. The entry and exit windows allow the neutron beam to pass the RFSR with less than 1% attenuation. Depolarization of the neutron beam in the aluminum is <~0.1%.

### B. Control electronics for RFSR operation

As described above the amplitude of the RF magnetic field generated by the solenoid is matched to the velocity of the neutrons to rotate the spin direction by π radians. Because the velocity of the neutrons is proportional to $1/t_{tof}$, the amplitude $B_1$ must be changed as $1/t_{tof}$ during each 50 ms long neutron pulse for π radians rotation. For 100-meV neutrons the required amplitude of the RF magnetic field is 0.38 mT, which is produced by a solenoid current of approximately 0.8 A. To reduce the voltage requirement for the amplifier that drives the solenoid ($L$=8.3 mH), a capacitor (C=3.6 pF) was added in series to produce a series resonant circuit with a resonant frequency approximately equal to the neutron spin precession frequency of 29 kHz for the 1-mT static field.

Figure 3 shows the circuit used to drive the solenoid. A frequency generator (FG2) produces a 29 kHz sine wave, which is amplified by a fixed-gain audio amplifier (A2). The output of the amplifier goes to a switchbox (SW) that selects whether the output is sent to the rotator coil or to a resistive dummy load, approximately equal to the impedance of the solenoid. The switch box also contains pickoff transformers to monitor current (X1) and voltage (X2) out of the amplifier. The outputs of these transformers are sampled at a rate of 62.5 kHz by 16 bit ADCs, allowing reconstruction of the waveforms. Typical reconstructed signals of the current sent to the solenoid (spin-on) or to the dummy load (spin-off) are shown in Fig. 4 left and right, respectively, and signals ramp up so that they can correct value by 4 ms.

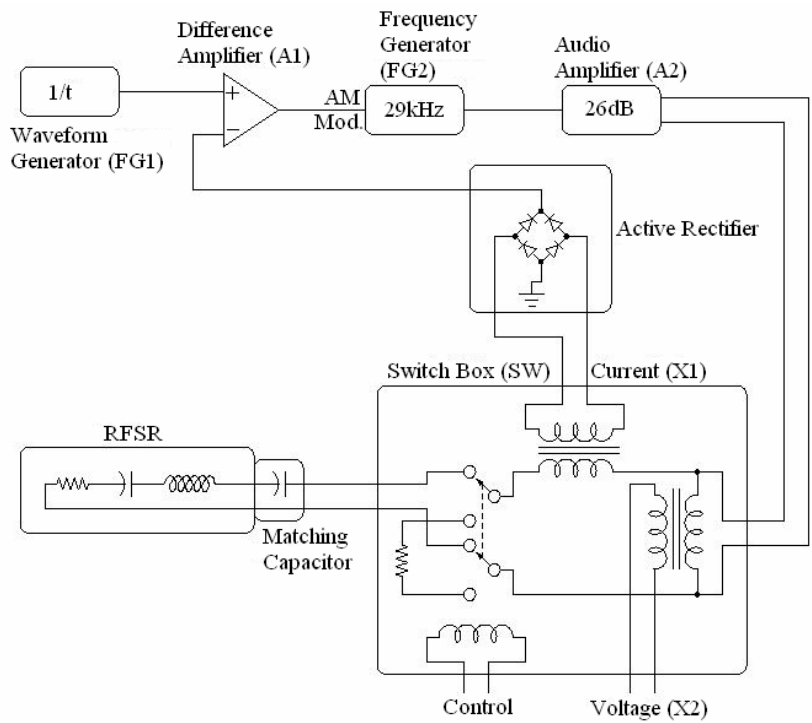

FIG 3. Schematic of the control electronic circuit of the RFSR. An arbitrary waveform generator (FG1) sends a $1/t$ ramp signal to a difference amplifier (A1). The amplifier outputs the signal difference between the $1/t_{tof}$ signal and the rectified 29 kHz feedback signal driven by the spin rotator coil. Current and voltage signals are measured by ADCs for each neutron pulse, see Fig. 4.

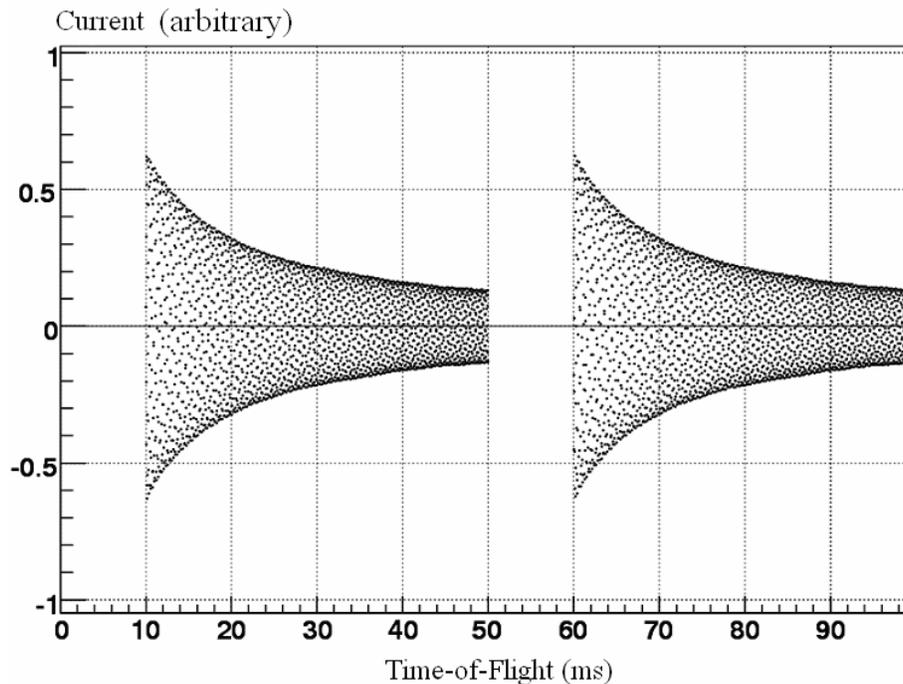

FIG 4. Typical reconstructed current-to-voltage signals for the solenoid (left) and for the resistive dummy load (right) as measured by ADC. The signals were sampled at the rate of 62.6 kHz over 40-ms long neutron frame. Note that the signals start at $t_{tof}$=0 ms but data acquisition starts at 10 ms.

The amplitude of the 29 kHz sine wave is varied to produce the required $1/t_{tof}$ behavior by amplitude modulating FG2. This amplitude modulation is produced by comparing an ideal ramp signal generated by FG1 to a full-wave rectified output of X1 using difference amplifier A1 which includes a single-pole filter with a 3-dB point of 1 kHz to filter out the 29 kHz component. This feedback mechanism maintains the accuracy of the amplitude ramp, compensating for gain and impedance drifts.

### C. Calculation of the RF fields

The RF field was calculated from a magnetic scalar potential $B(r,z) = -\nabla \phi(r,z)$. We approximate effects of the eddy currents in the aluminum cylinder by imposing the boundary conditions that the component of the magnetic field normal to the surface of the cylinder vanishes. Since the fields are cylindrically symmetric, the magnetic potential does not depend on the azimuthal angle.

Using the cylindrical coordinate system where the axis of symmetry is the $\hat{z}$-axis, we separate variables and seek solutions for radial and axial components of the potential in the form:

$$\phi(r,z) = f_n(r)g_n(z) \propto I_0(k_n r)\sin(k_n z) \propto K_0(k_n r)\cos(k_n z), \quad (4.1)$$

where $I_0(k_n r)$ and $K_0(k_n r)$ are modified Bessel functions. We divide the interior of the aluminum cylinder into two regions; one inside and the other outside of the solenoid. Because the magnetic field is regular on the axis, at $r = 0$, the coefficient of the $K_0(k_n r)$ term vanishes. Hence the coefficient of the $\cos(k_n r)$ term vanishes in both regions, the axial magnetic field must be an even function of $z$. With a boundary condition that at the ends of the cylinder $z = \pm L/2$ the $z$ component of the field vanishes, $B_z(r, z = \pm L/2) = 0$, we obtain that $k_n = (2n-1)\pi/L$. The magnetic potential can be written as a sum over $n$

$$\phi_{inner}(r,z) = \sum_{n=1}^{\infty} \alpha_n I_0(k_n r)\sin(k_n z),$$

$$\phi_{outer}(r,z) = \sum_{n=1}^{\infty} \sin(k_n z)[\beta_n I_0(k_n r) + \gamma_n K_0(k_n r)]. \quad (4.2)$$

Next, we impose two additional boundary conditions: $B_r(r = S, z)$ must be continuous and $B_r(r = R, z) = 0$, where $S$ is the radius of the solenoid and $R$ is the radius of the cylindrical aluminum shield. These two conditions determine $\beta_n$ and $\gamma_n$. The boundary condition $\frac{d}{dz}(\phi_{outer}(S,z) - \phi_{inner}(S,z)) = J$, where $J$ is the current density of the solenoid, determines $\alpha_n$. By using the orthogonality of the $g_n(z)$ we obtain

$$J \int_{-M/2}^{+M/2} \cos(k_n z)dz = k_n \int_{-L/2}^{+L/2} \cos^2(k_n z)dz[(\beta_n - \alpha_n)I_0(k_n S) + \gamma_n K_0(k_n S)],$$

and after integration

$$\frac{2J\sin(k_n M/2)}{k_n} = \frac{k_n L}{2}[(\beta_n - \alpha_n)I_0(k_n S) + \gamma_n K_0(k_n S)], \quad (4.3)$$

where $M$ is the length of the solenoid and $L$ the length of the aluminum shield. The calculated fields shown in Fig. 5, are a sum of the first 40 terms of the slowly converging series.

### D. Measurement of the RF fields

The field of the RFSR was measured to verify the results of the calculations. A small probe coil was fabricated and mounted on an arm so that the axis of the coil was either parallel or perpendicular to the axis of the RFSR. The probe was then stepped by a computerized scanner to map the RF field of the RFSR. The probe coil has 16 turns of enamel-coated Cu wire on a diameter of 7.8 mm. All the others components of the probe system were made of plastic. It was important to make sure that the axis of the probe coil was accurately perpendicular to the axis of the RF field since the radial field component $B_r$ is much smaller than the axial field component $B_z$ which could easily contaminate the measurement of the radial field component. For the same reason, the coil of the probe was carefully wound so that each turn was perpendicular to the axis of the probe coil.

A 100 mV$_{PP}$ sine wave, corresponding to 1 A current in the solenoid, at 29 kHz was connected to the reference input of a lock-in amplifier and to the input of the audio amplifier which was driving the series tuned $LC$ circuit of the RFSR. The signal from the probe coil was then connected to the input of the lock-in amplifier. The amplitude of the RF field was obtained from the root-mean-squared output voltage of the lock-in amplifier.

The $B_z$ and $B_r$ components at $r=0$ cm and 5 cm were mapped with 10-mm steps. The maximum of $B_z$ on axis was measured to be 0.42 mT at the center of the solenoid and increasing slowly with $r$. On the axis $B_r=0$ mT and off axis $B_r$ increases when approaching the ends of the solenoid. Neither field components were distorted much by the aluminum windows. In Fig. 5 we compare the measured $B_z$ at $r=0$ cm (○) and $r=5$ cm (×), and $B_r$ at $r=5$ cm (Δ) with the corresponding calculated fields (solid lines). They agree within 1%.

### V. SPIN REVERSAL EFFICIENCY OF THE RF SPIN ROTATOR

We have developed two independent methods to evaluate the neutron spin-flip efficiency of the RFSR over the phase space defined by the 1FP12 neutron guide. In the NPDGamma experiment the efficiency is one of the parameters required to extract the final physics γ-ray asymmetry

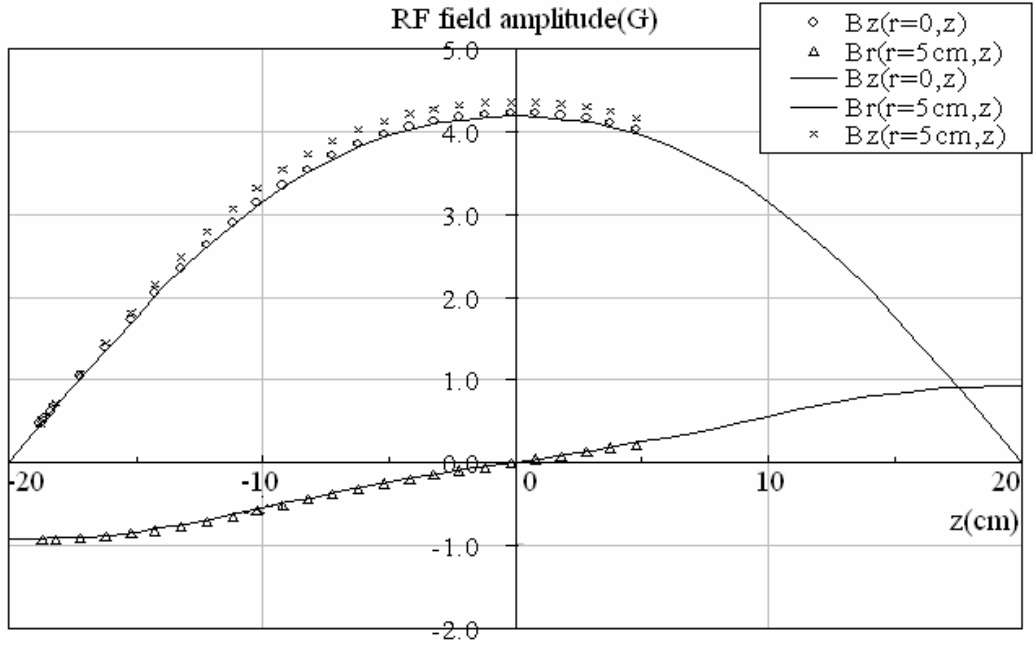

FIG 5. Comparison of measured (○, ×, and Δ) and calculated field components (solid lines) for $B_z(r=0$ cm$, z)$, $B_z(r=5$cm$, z)$, and $B_r(r=5$cm$, z)$ of the RFSR. The solenoid is 30 cm long and the RFSR (i.e. the RF field) is 40 cm long.

$$A_\gamma \equiv \frac{A_\gamma^{measured}}{P_n \cdot \varepsilon \cdot c}. \qquad (5.1)$$

Here $P_n$ is neutron beam polarization, $\varepsilon$ is the RFSR efficiency, and the constant $c$ contains the rest of the corrections for $A^{measured}_\gamma$ as discussed in Refs. [19,23].

    The spin-flip efficiency measurements consisted of a set of neutron transmission measurements with monitor M1, polarizer (P), M2, RFSR, analyzer (A), and M3 (see Fig. 1). The neutron beam was first polarized by the $^3$He spin filter with opacity $x_P = n_P \sigma l_P$ and $^3$He polarization of $P_P$. Here $n_P$ is the $^3$He number density, $l_P$ the length of the $^3$He, and $\sigma$ is the spin-dependent n-$^3$He absorption cross section for the reaction channel $J=0$. The cross section $\sigma$ has the well-known $1/v$-energy dependence, where $v$ is the neutron speed. After the spin filter the polarized neutrons enter the RFSR where the spin direction of beam was rotated by $\pi$ radians with efficiency $\varepsilon$, $0 \leq \varepsilon \leq 1$. If the RFSR was off, there was no change to the spin direction. After passing the RFSR, the polarization of the beam was analyzed by the polarization analyzer. After the polarization analyzer the transmitted neutrons were detected by monitor M3. The analyzer cell had opacity $x_A = n_A \sigma l_A$, where $n_A$ is the $^3$He number density, $l_A$ is the thickness of the $^3$He, and the $^3$He polarization in the cell is $P_A$.

    We used two different neutron transmission methods to measure the spin-flip efficiency. In the RFSR flip method the efficiency measurement was obtained by comparing the RFSR on- and off-transmissions. In the AFP $^3$He flip method, the transmission through the RFSR is measured before and after the neutron beam polarization is flipped by an AFP flip of the $^3$He polarization in the spin filter cell.

## A. Efficiency of the RFSR using the RFSR flip method

In the RFSR flip method, the polarizer and analyzer have fixed polarization directions and the spin-flip efficiency is obtained from the ratio of the transmission through the analyzer with the RFSR on and off. Let $x_A$ and $x_P$ be the opacities of the analyzer and polarizer and $P_A$ and $P_P$ the $^3$He polarizations of the analyzer and polarizer. The number of neutrons in the two spin states (+/- states corresponds to up/down states) incident on the analyzer when the RFSR is off is,

$$N_P^\pm = \frac{N_0}{2} T_P^\pm = \frac{N_0}{2} e^{-x_P(1 \mp P_P)}, \tag{5.2}$$

assuming no loss of neutrons after the polarizer. Here $N_0$ is the total number of un-polarized neutrons entering the polarizer. The total number of neutrons in the two spin states, arriving at the analyzer cell when the RFSR is off, is then

$$N_P = N_P^+ + N_P^- = N_0 e^{-x_P} \cosh(x_P P_P).$$

The number of the transmitted neutrons through the analyzer cell is,

$$N_{\text{off-A}}^\pm = N_P^\pm e^{-x_A(1 \mp P_A)} \tag{5.3}$$

and thus the total number of neutrons after the analyzer when the RFSR is off, is

$$N_{\text{off-A}} = N_{\text{off-A}}^+ + N_{\text{off-A}}^- = N_0 e^{-(x_A + x_P)} \cosh(x_A P_A + x_P P_P). \tag{5.4}$$

When the RFSR is on, the beam polarization ($P_n$) rotates $\pi$ radians about the $\hat{z}$-axis while the neutrons traverse through the RFSR. The spin-reversal is not perfect and we define the spin rotation efficiency $\varepsilon$,

$$P_{\text{n-on}} \equiv -\varepsilon P_{\text{n-off}}, \tag{5.5}$$

where $0 \leq \varepsilon \leq 1$ and $P_{\text{n-off}} = \tanh(x_P P_P)$. After the RFSR the number of neutron in the two spin states are

$$N_{\text{P-on}}^+ = (1-\varepsilon)\frac{N_0}{2} e^{-x_P(1-P_P)} + \varepsilon \frac{N_0}{2} e^{-x_P(1+P_P)}$$

$$N_{\text{P-on}}^- = (1-\varepsilon)\frac{N_0}{2} e^{-x_P(1+P_P)} + \varepsilon \frac{N_0}{2} e^{-x_P(1-P_P)}. \tag{5.6}$$

The transmission of the neutrons of the known spin states through the analyzer cell is

$$N_{\text{on-A}}^+ = N_{\text{P-on}}^+ e^{-x_A(1-P_A)}$$

$$N_{\text{on-A}}^- = N_{\text{P-on}}^+ e^{-x_A(1+P_A)}. \tag{5.7}$$

The total number of neutrons which pass through the analyzer is,

$$N_{\text{on-A}} = N_0 e^{-(x_A + x_P)}\left[(1-\varepsilon)\cosh(x_A P_A)\cosh(x_P P_P) + \varepsilon \cosh(x_P P_P - x_A P_A)\right]. \tag{5.8}$$

The transmitted neutrons through the analyzer when the RFSR is off and on, are given by Eqs (5.4) and (5.8). Let $M3_{\text{off(on)}}$ be the monitor signals from neutron transmission through the analyzer with the RFSR on and off. The efficiency of the RFSR, $\varepsilon$, can be now extracted from the measured transmission ratio

$$\frac{M3_{on}}{M3_{off}} = \frac{(1-\varepsilon)\cosh(x_A P_A)\cosh(x_P P_P) + \varepsilon \cosh(x_A P_A - x_P P_P)}{\cosh(x_A P_A + x_P P_P)}, \quad (5.9)$$

assuming that parameters $x_A$, $x_P$, $P_A$, and $P_P$ are known.

## B. Efficiency of the RFSR using the AFP $^3$He flip method

A nearly-perfect reversal of neutron beam polarization can be achieved by performing a $^3$He polarization reversal in the polarizer by AFP. After an AFP flip the analyzed beam polarization can be compared to the RFSR flipped beam polarization to determine the beam polarization reversal efficiency $\varepsilon$. To derive the number of neutrons entering the monitor M3 after the AFP, incident on M3, we need only to change the sign of the $^3$He polarization $P_P$ in the polarizer, in Eq. (5.2)

$$N_{AFP}^{\pm} = \frac{N_0}{2} T_P^{\pm} = \frac{N_0}{2} e^{-x_P(1\pm P_P)} \quad (5.10)$$

and then transmit these neutrons through analyzer and compare the result with the result from the RFSR flip given by Eq. (5.8). The total number of the AFP flipped neutrons incident on M3 is

$$N_{AFP} = N_{AFP}^{+} + N_{AFP}^{-} = N_0 e^{-(x_A + x_P)} \cosh(x_A P_A - x_P P_P). \quad (5.11)$$

The RFSR efficiency can be then extracted from the measured transmission signals $M3_{off}$, $M3_{on}$, and $M3_{AFP}$;

$$\frac{N_{off} - N_{on}}{N_{off} - N_{AFP}} = \frac{M3_{off} - M3_{on}}{M3_{off} - M3_{AFP}} = \frac{1+\varepsilon}{2}. \quad (5.12)$$

## C. Measurement, analysis, and results of the RFSR efficiency

In Eqs. (5.9) and (5.12), the determination of the spin-flip efficiency using the polarizer and the analyzer cells depends on $^3$He polarization and $^3$He thickness in the cells seen by neutron beam through the parameters $x_A$, $x_P$, $P_A$, and $P_P$.

All of these parameters were constant during the measurements except for $P_A$, which varied slowly with time. Here we describe the determination of these parameters. The $^3$He in the polarizer was polarized by continuous online spin-exchange optical pumping on the neutron beam. The polarization was measured to be 46±2% by neutron transmission. The description of the $^3$He polarization and its performance can be found in Ref. [17]. The neutron polarization analyzer cell was a 2.5 cm diameter sphere polarized outside the experiment using a single bar diode laser with a spectrum narrowed by an external cavity [21]. The $^3$He polarization of the analyzer was first measured off-line by NMR separately for each spin flipper efficiency measurement with the maximum polarization of 61±5%. The analyzer cell was then transported to the experiment in a few gauss magnetic field to minimize depolarization: 1-2% polarization was typically lost. The decay time constant of the polarization in the room temperature cell was measured to be 130 hours. After the neutron spin-flip efficiency measurements, the cell was returned to the spin-exchange optical pumping system and the polarization was remeasured. The NMR results of the $^3$He polarization were also verified by neutron transmission measurements. The largest source of systematic error in determining the $^3$He polarization of the analyzer cell by neutron transmission was uncertainty in the effective $^3$He thickness from the curved GE180 glass walls of this spherical cell. The systematic uncertainty involved using the $^3$He polarizer is discussed in Ref. [17].

In the analysis each data set consisted of signal runs and their corresponding background runs taken with the beam shutter closed. A single run contained 1000 neutron pulses. The monitor M3 data was normalized with the M1 data and then background runs were subtracted. The efficiencies were calculated by integrating the monitor signals over the neutron energy range of 3-18 meV. The lower energy side of this range was defined by an upstream neutron chopper that started closing off the beam at 30 ms for background measurements from 34 ms to 50 ms. Figure 6 shows the measured neutron spectra from M1, M2, and M3 as a function of neutron time of flight, with M2 and M3 multiplied by a factor of three and two, respectively. The 1/$v$-dependence of the $^3$He neutron absorption cross section preferentially absorbs the low energy neutrons. Figure 7 shows M3 signals as a function of time of flight for eight consecutive spin states ↓↑↑↓↑↓↓↑ where ↓ (↑) represents the RFSR off-state (on-state).

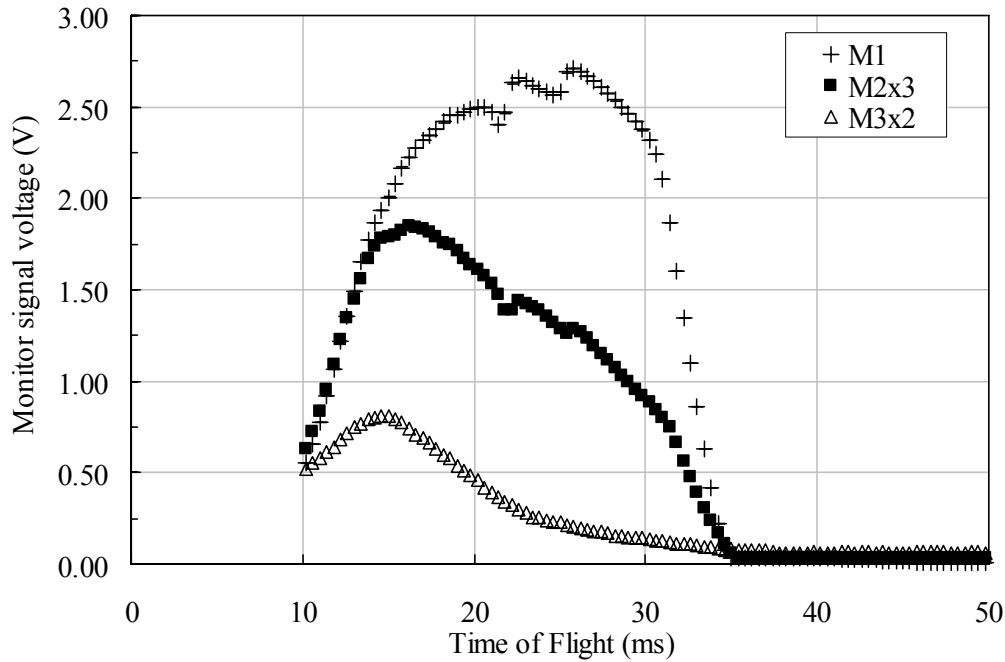

FIG 6. Signals from the monitors M1, M2, and M3 as a function of neutron time-of-flight. M1 and M2 have the same preamplifier gains whereas M3 has about 50 times larger gain. The changes in the neutron energy spectrum at the different monitor positions are mainly caused by the $^3$He absorption and its 1/$v$-dependence. The signals are shifted in time so that they can be compared easily.

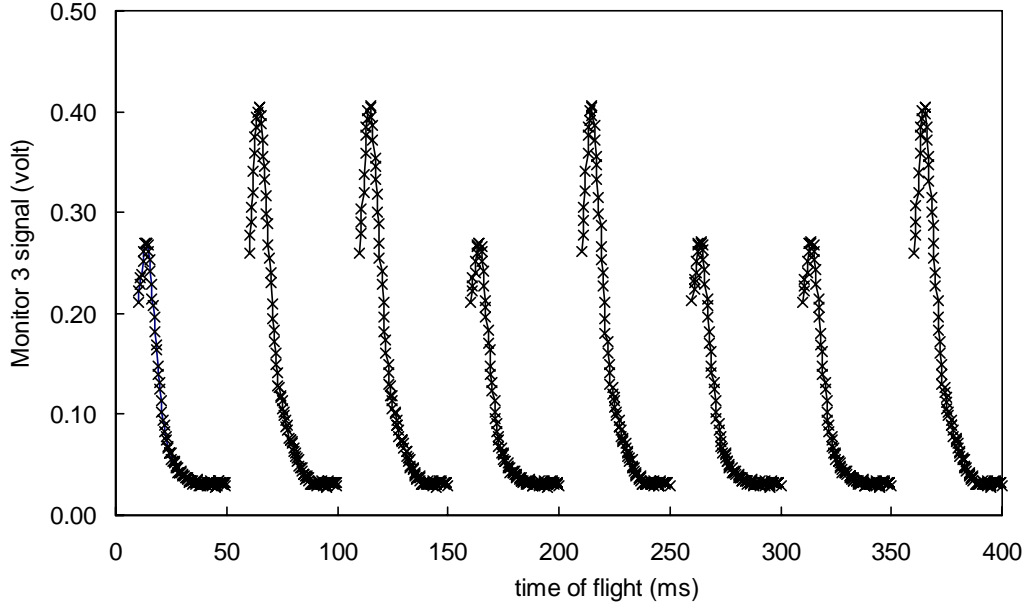

FIG 7. Eight consecutive neutron pulses corresponding to the eight-step spin sequence ($\downarrow\uparrow\uparrow\downarrow\uparrow\downarrow\downarrow\uparrow$) of the RFSR measured by M3 as a function of time of flight. Each neutron time of flight frame is 50 ms long. Beam absorption in the $^3$He analyzer cell is much larger when the beam polarization is opposite to the $^3$He polarization.

Figure 8 shows two superimposed measured efficiencies as a function of the relative RF field change, $\Delta B_1/B_1$. Symbols ($\Delta$) represent data measured using the RFSR flip method of Eq. (5.9) and the data points ($\bullet$) are measured with the AFP $^3$He flip method of Eq. (5.12). In the measurement the static field is set to correspond to the maximum RFSR efficiency value i.e. to resonance where $\varepsilon=0$ in Eq. (2.1). The lines are fits to data using quadratic functions. The both methods give the same maximum efficiency in errors.

Figure 9 presents measured efficiencies as a function of a relative change in the static field, $\Delta B_0/B_0$ when $B_1$ is kept fixed at the maximum RFSR efficiency. The data set ($\Delta$) is measured using the RFSR flip method and the data set ($\bullet$) using the AFP $^3$He flip method. Each data set is fitted to a quadratic function.

In order to determine the spin-flip efficiency over the entire phase space of the beam, the RF field amplitude scan was performed on three off-axis analyzer positions: one in the vertical position ($x=0$, $y=3.3$ cm) and two in the horizontal positions ($x=\pm3.3$ cm, $y=0$), and then efficiencies were measured using the RFSR-flip method. Results are presented in Fig. 10. The three data sets are fitted to quadratic functions: the line (- - -) corresponds to $x=+3.3$ cm, the line ($- - -$) to $x=-3$ cm, and the line (———) to $y=+3$ cm.

Near maximum the values of the $x=+3.3$ cm position seem systematically lower than the values of the other positions. The maximum efficiency values given by the fits have been used for determination of the overall efficiency across the beam.

From the results of the on-axis (Fig. 8) and the off-axis (Fig. 10) measurements, the overall spin–flip efficiency of 98.0±0.8% was obtained across the beam. Uncertainty in the figures is statistical.

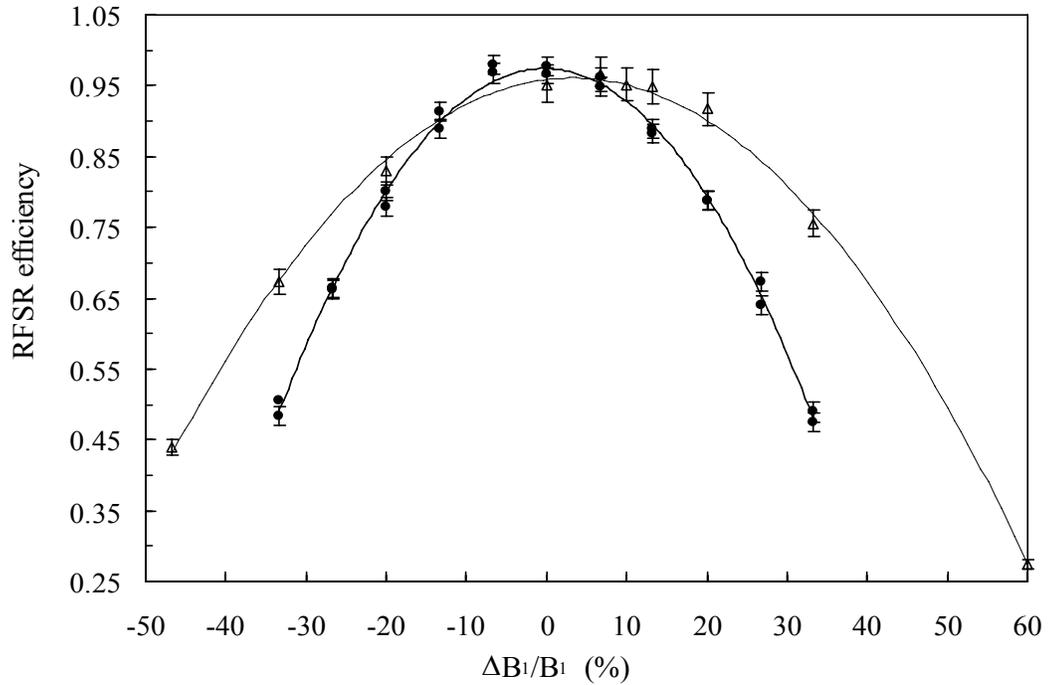

FIG 8. The measured efficiencies of the RFSR using the RFSR flip method (Δ) and the AFP $^3$He flip method (•) are plotted as a function of a relative change in the RF-field, $\Delta B_1/B_1$. The dashed and solid curves are quadratic fits to data.

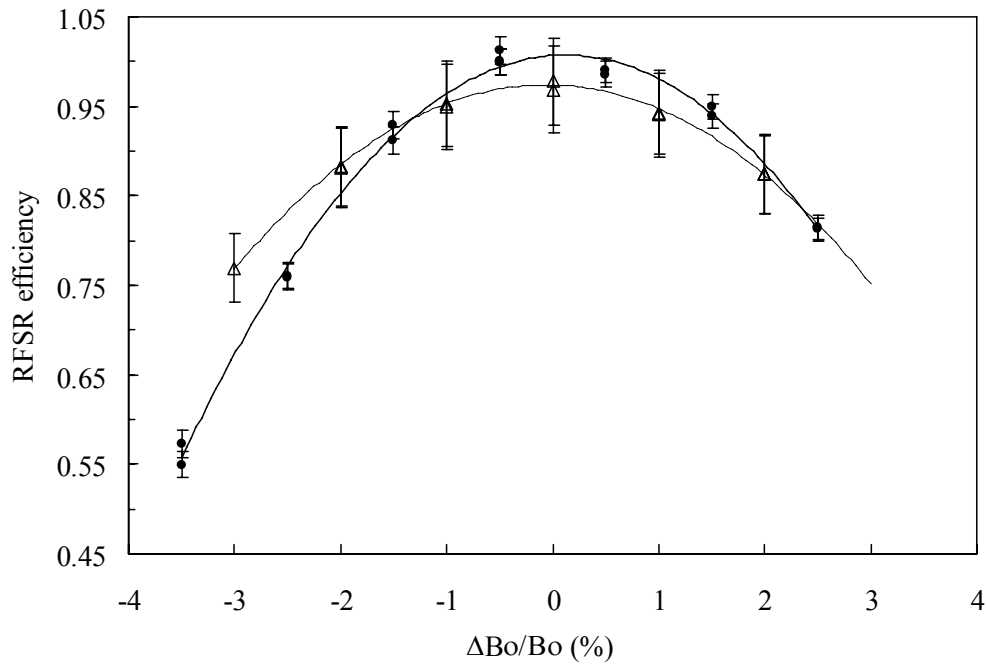

FIG 9. Measured efficiencies of the RFSR are shown in a function of a relative change in the static magnetic field. Triangles are measured with the RFSR flip method and solid points with the AFP $^3$He flip method. Lines are results of quadratic fits to the data set.

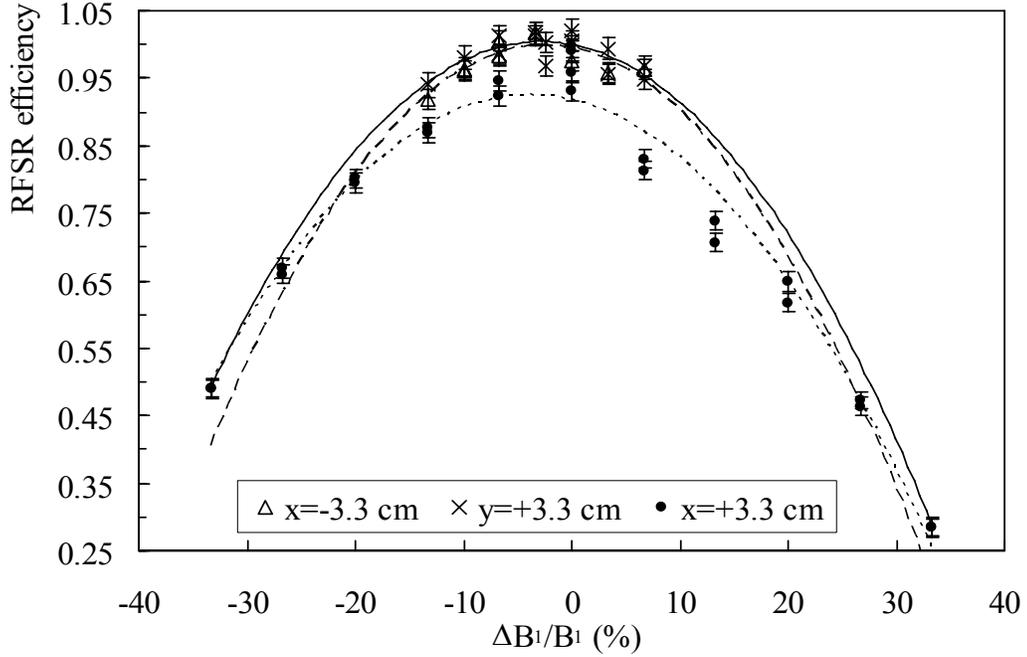

FIG 10. The spin-flip efficiency of the RFSR on off-axis at $x=\pm 3.3$ cm (circle and triangle) and y=+3.3 cm (cross) as a function of the RF field amplitude $B_1$. The data are measured using the RFSR flip method. The lines (- - - -, – – –, and ——) are quadratic fits to the data set.

In the RFSR-flip method the main systematic uncertainty is caused by the accuracy of the $^3$He thickness and $^3$He polarization; $x_A$ and $P_A$ for the analyzer and $x_P$ and $P_P$ for the polarizer. These parameters were determined by performing auxiliary neutron transmission measurements. The relative uncertainties in $x_A$ and $P_A$ are 0.02 and 0.04, and in $x_P$ and $P_P$ are 0.02 and 0.03, respectively. The length of the flight path was determined by using the time-of-flight locations of known beryllium Bragg edges, where the total neutron cross section in polycrystalline aluminum changes discontinuously as a function of neutron energy. The relative error in the flight path length is $2\times 10^{-3}$ [20]. A typical error from the fittings is 0.01. The overall systematic error in the efficiency when using the RFSR flip method, is under 5%. In the AFP $^3$He flip method where the knowledge of the polarization and the thickness of the $^3$He are not required, the systematic error is < 2%.

Due to thermalization processes in the moderator, geometric effects like the time distribution of neutrons from the neutron moderator, path length differences, neutron beam divergence, and small changes in the kinetic energy of the neutrons by the RFSR for off-resonance neutrons, there is a distribution of spin rotation angles around $\pi$ radians in the RFSR. The fractional width of this distribution is estimated to be at the ~1% level. This departure of the spin-flip probability from unity translates into a ~$10^{-4}$ effect in the spin-flip efficiency.

## D. Efficiency of the RFSR with parity-violating γ-ray asymmetry on $^{35}$Cl

As an independent verification of the performance of the RFSR, we measured the parity-violating γ-ray asymmetry $A_\gamma$ from polarized neutron capture on $^{35}$Cl using our current-mode CsI gamma detector array, described in [19] as a function of the relative change in the RF field amplitude $B_1$. At low neutron energies, the γ-ray asymmetry on $^{35}$Cl is large and well known [22].

Using parity-violating $A_\gamma$ as a beam polarization analyzer, the efficiency ε can be calculated from Eq. (5.1) if beam polarization $P_n$ and the instrumental constant $c$ are known. To reach a 20% precision in the spin-flip efficiency measurement with the Cl-target a few hours beam on target was required. Figure 11 shows a result of such a measurement. The results in Figs. 8 and 12 should be about the same, however a small geometrical difference between these two measurements is that the Cl measurement was performed with 5 cm beam collimation. The systematic errors of the Cl measurement where γ-rays from neutron capture on $^{35}$Cl are detected, are completely different compared to the neutron transmission measurements with the $^3$He analyzer.

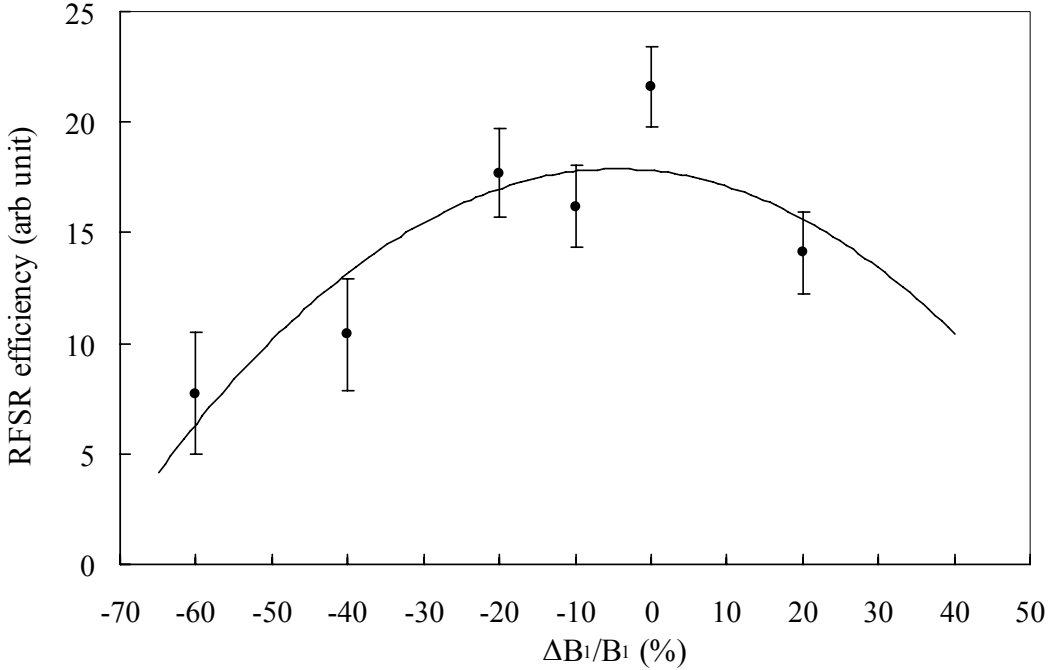

FIG 11. The efficiency of the RFSR is shown as a function of a relative change in the RF field $\Delta B_1/B_1$. The efficiency measurement with cold neutrons is based on the well-known parity-violating γ-ray asymmetry on $^{35}$Cl (see text). The solid line is a fit to a quadratic function of Eq. (2.1) the spin-flip efficiency.

## VI. SYSTEMATIC EFFECTS PRODUCED BY THE RFSR ON THE NPDGAMMA EXPERIMENT

Since the NPDGamma experiment aims to measure the γ-asymmetry with a precision of $1\times10^{-8}$, false asymmetries have to be controlled to an even greater accuracy. Here we consider systematic effects caused by the RFSR on the measured $A_\gamma$ in the n-p capture reaction at cold neutron energies. To produce a false asymmetry, a source has to produce a signal which is in phase with the state of the neutron spin. Otherwise it will average out with time and appear as another source of random noise. The origin of these systematic effects can be either instrumental or due to neutron interactions with matter other than n-p capture in the liquid parahydrogen target. The physics origins of the systematic effects have been discussed in detail in Ref. [9,24] and will not be repeated here. Since the parity-violating asymmetry $A_\gamma$ is energy independent over the range of neutron energies in this experiment, systematic effects, which in many cases have energy dependence [9,24], can be isolated using neutron time-of-flight.

The most serious potential source of an instrumental systematic effect is the RF field of the spin rotator. It is perfectly coherent with the neutron spin state and there is an electronic difference in the RFSR between the spin-up and spin-down states. To minimize any potential problem from electronic pickup, the RF power is held constant and switched either to the RFSR or to a dummy load (see section IV.C).

### A. Raw parity-violating γ-ray asymmetry

The 48 CsI(Tl) crystals of the γ-ray detector array are symmetrically mounted around the target. The detector array is divided into symmetric upper and lower detector pairs [19]. This allows the experiment to measure $A_\gamma$ in single neutron pulse (in single spin state)

$$A_\gamma^{\text{measured}} = \frac{U_\uparrow - D_\uparrow}{U_\uparrow + D_\uparrow},$$

where $U_\uparrow(D_\uparrow)$ is the sum of the signal of the upper (lower) detector pairs when the neutron spin is up. The γ-ray asymmetry $A_\gamma$ can be determined from $A_\gamma^{\text{measured}}$ after applying corrections for beam polarization, RFSR spin-flip efficiency, neutron depolarization in target, and detector acceptance, see Eq. (5.1). Since the gains of the upper and lower detector pairs are difficult to equalize accurately enough, we switch between the upper and lower detector pairs by reversing the neutron spin direction by the RFSR. The raw asymmetry is thus a difference between two spin states [18,23],

$$A_\gamma^{\text{measured}} = \frac{U_\uparrow - D_\uparrow - (U_\downarrow - D_\downarrow)}{U_\uparrow + D_\uparrow + U_\downarrow + D_\downarrow}.$$

### B. Additive and multiplicative coupling of RF power of the RFSR to the detector system

A design requirement for the NPDGamma experiment is that it must be possible to measure any electronic effects resulting in a false asymmetry to the same accuracy desired for $A_\gamma$ in a time short compared to the measurement time of the experiment. Instrumental systematic effects caused by the RFSR may introduce false asymmetries have been extensively tested for and are described in Ref. [19].

An additive effect, in which a false asymmetry may be induced as a result of an additional signal in the detector electronics, can be produced by pickup of the spin rotator signal. The additive effects have been searched for by measuring $A_\gamma$ without the incident neutron beam but operating the RFSR. Due to a small width of the distribution of the electrical noise, we can measure $A_\gamma$ in a few hours to an accuracy of $10^{-9}$. The noise observed in the detectors, with the spin rotator running, did not significantly change from the measured and calculated detector noise in the 48 detectors, which includes contributions from dark currents and cosmic-ray background [19,20].

A multiplicative effect is a gain change in the detector system, caused, for example, by such as a spin rotator magnetic field leakage into the detector system and a consequent gain change. A search for multiplicative effects is accomplished by using the two LEDs per crystal to simulate light intensities expected to be produced by neutrons during an actual measurement.

For the $10^{-9}$ accuracy in $A_\gamma$ we needed 3-4 times more time to perform a search of multiplicative effects compared to the additive effects. The reason is the increased width of fluctuations of the LED introduced light to the noise spectrum.

These measurements have shown that the shielding of the RFSR is sufficient to isolate electronically the RFSR power from the rest of the experiment in the sensitivity level of $10^{-9}$ [19].

### C. Control of drifts by an eight-step spin sequence

To control systematic effects caused by time-dependent drifts in detector efficiencies, we use a symmetric eight-step spin sequence, (+--+-++-) or its complement. The four spin-up states are added together to obtain $U_\uparrow$ and $D_\uparrow$ and then other four to obtain $U_\downarrow$ and $D_\downarrow$. The asymmetry $A_\gamma^{\text{measured}}$ is calculated for each eight-state sequence. Use of signals ($S$) of such spin-state octets cancels both linear drifts (e.g. $+S-2S-3S+4S-5S+6S+7S-8S=0$) and quadratic drifts (e.g. $+S-4S-9S+16S-25S+36S+49S-64S=0$) over the octet.

### D. Other systematic effects on the NPDGamma produced by the RFSR

Two additional sources of systematic effects are considered. One is a change in the kinetic energy of a neutron caused by the RFSR. As discussed in section II this is possible only when the static field is different in the entry of the RFSR compared to the exit. We know from the field measurement that the field difference is less than 0.2 µT across of the RFSR, then the relative change in the kinetic energy of a 4-meV neutron is less than $10^{-5}$ leading to the second order effects on $A_\gamma$ which are less than $10^{-10}$.

To estimate the effect of the beam up and down motion by the Stern-Gerlach steering, we used the measured result for the field gradient $\frac{\partial B_y}{\partial z} \leq 10$ µT/m between the RFSR and the end of the LH$_2$ target. The small difference between the beam positions in the detector with the neutron spin up- and down-state, leads to a 10$^{-10}$ change in the solid angle of the detector.

## VII. Conclusion

We have developed a RF resonant spin rotator to reverse the neutron polarization in a pulsed cold polarized neutron beam. The efficiency of the spin rotator was measured to be 98.0±0.8% for neutron energies from 3.3 to 18.4 meV over the full phase space of the beam, which possesses a 9.5 cm × 9.5 cm cross sectional area and a transverse momentum spread set by a 15 m long *m*=3 supermirror neutron guide as described and characterized in Ref. [14]. In addition, we have placed upper bounds on the size of the systematic effects that the RF spin rotator introduces into the NPDGamma experiment. The spin rotator has already been used in a number of measurements of parity violation in polarized neutron capture in several nuclei [23].

We note that many of the properties of both the resonant RF spin rotator itself and the techniques used to characterize its efficiency are also applicable to other polarized neutron measurements using pulsed cold neutron beams. Since the rotator operates in a uniform static magnetic field, it is fully compatible with the small magnetic field gradient requirements of polarized $^3$He neutron spin filters. As the performance of $^3$He neutron spin filters continues to improve [17] their use in neutron scattering spectrometers promises to expand, and it is possible that their broad phase space acceptance and wide dynamic range in neutron energy can make possible new types of neutron scattering spectrometers. For many such instruments of the future it is likely that a resonant RF spin rotator will be an attractive choice for neutron spin reversal. The combined use of polarized $^3$He neutron polarizers and analyzers and RF spin rotators is also foreseen in future polarized neutron β-decay experiments at pulsed neutron sources [25], where the unique properties of $^3$He neutron spin filters are well-suited to absolute determination of the neutron beam polarization. For the latter application further characterization of the systematic effects is underway.

## ACKNOWLEDGEMENTS

The authors would like to thank Mr. G. Peralta for his technical contribution to the success of this work. The work was supported in part by the U.S. Department of Energy (Office of Energy Research, under Contract W-7405-ENG-36), the National Science Foundation (Grants No. PHY-0100348 and PHY-0457219), the Natural Science and Engineering Research Council of Canada, and the Japanese Grant-in-Aid for Scientific Research A12304014. M. Snow acknowledges the Institute for Nuclear Theory at the University of Washington for partial support during the composition of this paper.